\documentclass[usenatbib]{mn2e}
\usepackage{graphicx}

\begin{document}

\title[Cosmic Evolution of Metallicity]{The Cosmic Evolution of Metallicity from the SDSS Fossil Record}

\author[Panter et al.]{Benjamin Panter$^{1}$, Raul Jimenez$^{2}$, Alan F. Heavens$^{1}$, Stephane Charlot$^{3}$\\
$^1$Institute for Astronomy, SUPA, University of Edinburgh, Royal Observatory, Edinburgh EH9-3HJ, UK; bdp, afh@roe.ac.uk \\
 $^2$Institute of Space Sciences(ICE), CSIC-IEEC/ICREA, UAB campus, Barcelona, Spain;
jimenez@ieec.uab.es\\
 $^3$Institut d'Astrophysique de Paris, UMR 7095, 98 bis Boulevard Arago, F-75014 Paris, France;
charlot@iap.fr}

\maketitle

\begin{abstract}
We present the time evolution of the stellar metallicity for SDSS
galaxies, a sample that spans five orders of magnitude in stellar
mass ($10^7 - 10^{12}$ M$_{\odot}$). Assuming the \citet{BC03}
stellar population models, we find that more massive galaxies are
more metal-rich than less massive ones at all redshifts; the
mass-metallicity relation is imprinted in galaxies from the epoch of
formation. For galaxies with present stellar masses $ > 10^{10}$
M$_{\odot}$, the time evolution of stellar metallicity is very weak,
with at most $0.2-0.3$ dex over a Hubble time- for this reason the
mass-metallicity relation evolves little with redshift. However, for
galaxies with present stellar masses $ < 10^{10}$ M$_{\odot}$, the
evolution is significant, with metallicity increasing by more than a
decade from redshift 3 to the present. By being able to recover the
metallicity history, we have managed to identify the origin of a
recent discrepancy between the metallicity recovered from nebular
lines and absorption lines. As expected, we show that the young
population dominates the former while the old population the latter.
We have investigated the dependence on the stellar models used and
find that older stellar population synthesis codes do not produce a
clear result. Finally, we have explored the relationship between
cluster environment and metallicity, and find a strong correlation
in the sense that galaxies in high density regions have high
metallicity.
\end{abstract}


\section{Introduction}

The quality of spectra of the observed light of unresolved stellar
populations has reached a sufficient accuracy that it is now
possible to make detailed studies of the physical properties of the
stellar populations in these galaxies. While much attention has
focussed on the time evolution of the star formation rate of
galaxies \citep{PHJ03,HPJD04,PHJ04,Sodre05,Mathis06,Ocvirk06}, a
closely related quantity, the evolution of the metallicity of the
star forming gas in galaxies, has not received as much attention \citep{PHJ03,Erb06,Bouchet07,Cid07,Erb08,Gallazi08}.
One reason is that while it is relatively straightforward to measure
the star formation rate from massive stars at different redshifts,
it is not easy to determine its metallicity, which requires
high-quality spectra. Further, in order to cover a large range in
mass and lookback time one would need to observe a very large area of
the sky, of the order of thousands of square degrees.

One alternative approach is to use the fossil record in the local
universe and reconstruct the star and metallicity history from this
record. This however has its own limitations \citep{Ocvirk06,PJHC06,
VESPA}. Using MOPED, a tool that allows rapid extraction of the
physical parameters of the stellar populations of galaxies from
their spectra, our group has shown how it is possible to use the
fossil record to obtain information about the stellar populations of
galaxies at large lookback times \citep{PJHC06}. In this work we
concentrate on what can be learned about the metallicity
of those stellar populations.

The evolution of metallicity of the stellar populations of galaxies
as a function of redshift is highly relevant as it can tell us about
how the interstellar medium is being enriched, potentially what the
initial mass function is (provided the theoretical yields of stars
are well known) and how this enrichment depends on environment
\citep{SJPH06} or mass, star formation history and  age of the
galaxy \citep{PJHC06}.

Determining the  metallicity history of galaxies has indeed a long
history. The first attempt is to use our own galaxy as
representative and measure metallicities in its
individual stars or stellar clusters (see, for example, the review
by \citet{FB06}). There are two major advantages:  one can obtain high
S/N detailed spectra, and the stellar populations are individually
resolved. The main disadvantage is that our galaxy is not typical of
others in the Universe.

The second route is to determine the metallicity of galaxies from
the spectrum of their integrated light. The pioneering grid of
models by \citet{W94} allowed this for simple stellar populations,
i.e. elliptical galaxies. His models provided Lick equivalent widths
in the optical region that were sensitive to the overall metallicity
of the stellar population. Indeed the application of this technique
has allowed many researchers to determine the metallicity of
galaxies in the nearby (e.g. \citet{Gallazzi06}) and distant (e.g.
\citet{Pettini01}) universe. The ratios of various emission lines
can also be used to determine the metallicity of the gas in a
galaxy, although careful calibration is required \citep{Kewley02,
Tremonti04,Erb06}.

However, \cite{HPJD04} have shown that for an ensemble of galaxies
it is possible to go beyond simply recovering the mass-weighted
metallicity of the stars and recover the metallicity history in many
time bins. This is not necessarily true for individual galaxies
(e.g. \citet{Ocvirk06,VESPA}), which may have noisy recoveries, but has been shown to be unbiased
when averaged over a sufficiently large sample. The recovery of the
star formation history and metallicity for galaxies is similar to
the ``population boxes'' for nearby resolved stellar populations in
local group galaxies (e.g. \citet{Gallart05}).

In this paper we recover the metallicity history of SDSS galaxies
from their spectra. We determine its evolution as a function of
mass, how the mass-metallicity relation depends on the age of the
stellar population and how sensitive our findings are to the stellar
population models used. The paper is organized as follows: in \S 2
we describe the SDSS sample used and in \S 3 we describe briefly the
method to determine metallicities. In \S 4 we show the derived
cosmic metallicity history and in \S 5 we present a map of the
stellar metallicity history. \S 6 and 7 concern the mass-metallicity
relation and how it evolves with redshift and mass. \S 8 shows the
impact of the choice of stellar population model on our findings. We
conclude with a general discussion of our findings in \S 9.

\section{The sample}

As described in \citet{PJHC06}, the MOPED algorithm has been used to
extract star formation and metallicity histories of a magnitude
limited sample of about $300,000$
galaxies drawn from the Third Data Release of the Sloan Digital Sky
Survey (SDSS DR3; \citet{SDSS-DR3}).

Our main galaxy sample is determined by red apparent magnitude
limits of $15.0 \le m_r \le 17.77$, and we also place a cut on
surface brightness of $\mu_r<23.0$. The magnitude limits are set by
the SDSS target selection criteria, as discussed in
\citet{SDSS-DR3}. The target criteria for surface brightness was
$\mu_r<24.5$, although for $\mu_r>23.0$ galaxies are included only
in certain atmospheric conditions. In order to remove any bias we
have therefore cut our sample at $\mu_r<23.0$. At low redshifts the
Sloan galaxies are subject to shredding - where a nearby large
galaxy is split by the target selection algorithm into several
smaller sources. To reduce this effect, for our star formation
analysis we use a range of $0.005<z<0.34$. We want to derive
properties that avoid dependence on the SFH of the galaxies, and
hence wish to avoid inverse $V_{max}$ weighting where possible. We therefore
choose mass cuts around a narrow redshift range of 0.1 for the
deriving the cosmic metallicity evolution, and show colors
normalised to the total number of galaxies in that mass range in the
mass-metallicity diagrams.

\section{Determining the metallicities of galaxies}

In order to determine the optimal multi-population fit for each
spectrum it is necessary to allow as free a star formation history
as possible. A parametrization of 11 SSPs spaced logarithmically in
lookback time, each with independent and variable metallicity, was
allowed along with a one parameter slab extinction based on the
\citet{Gordon03} LMC curve. In order to fully assess the resulting
23 dimensional likelihood surface it was necessary to use the
MOPED\footnote{Massively Optimized Parameter Estimation and Data
Compression}\citep{HJL00, Reichardt01} algorithm coupled with a
Markov Chain Monte Carlo search method \citep{PHJ03}, developed to
use high spectral resolution models \citep{PJHC06}. For each galaxy
the best solutions from a series of $100$ randomly-seeded conjugate
gradient searches were used to seed a Markov chain of $10^6$ steps.
The individual search traces were examined to confirm adequate
exploration of the parameter space and estimate solution
convergence.

In our approach, for each galaxy we describe its star formation by
11 lookback time bins (spaced logarithmically) and for each of these we determine
the mass fraction transformed into stars and its metallicity. The
first interesting result, before looking at the metallicity history
of galaxies, is to look at the overall metallicity of the stars in
SDSS galaxies today. It is important to realize that there must be
some weighting of the metallicity values recovered - the metallicity
of a population that contributes very little light to the final
spectrum is not constrained, so should be discounted. Since the
MOPED method recovers the complete star formation history of a
galaxy it is possible to calculate the average metallicity with
different weights. Traditional lick index analysis considers the
light-weighted metallicity, but with knowledge of the SFH and IMF it
is possible to determine present mass fraction (PMF) or original
star formation fraction (SFF)\footnote{i.e., the fraction of total
star formation over the galaxy's history} average metallicities
weighted by either.

\begin{figure}
\includegraphics[width=\columnwidth]{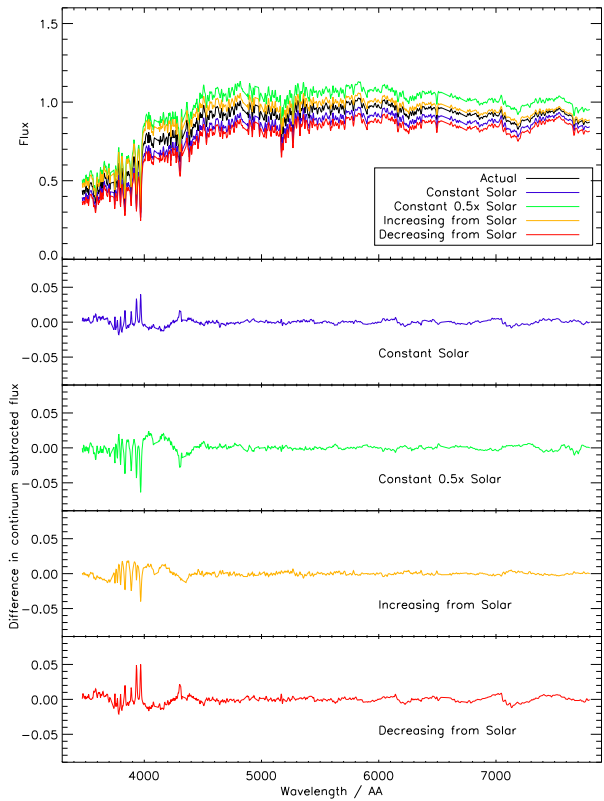}
\caption{A synthetic spectrum corresponding to the mean star
formation history of SDSS galaxies, and the effect of changing the
metallicity history while maintaining the same star formation
history. The lower panels show the difference between the
continuum-subtracted mean spectrum and a model produced using the
different continuum-subtracted alternate metallicity models,
illustrating that changes in metallicity history can be seen in both
the low and high frequency spectral signal. In this figure,
increasing from solar refers to an initial metallicity of $Z_\odot$
in the oldest bin rising to $2.5Z_\odot$ in the youngest and
decreasing from solar an initial metallicity of $Z_\odot$ falling to
$0.02Z_\odot$. } \label{fig:compo}
\end{figure}

In Fig.~\ref{fig:compo} we illustrate how a synthetic galaxy
spectrum responds to changes in metallicity history, by comparing a
synthetic spectrum created using the mean star and metallicity
history of SDSS galaxies recovered by MOPED with models that have
the same star formation history but different metallicity histories.
It is clear that changing the metallicity has a significant effect
on both the continuum level and depth of absorption features.

We can distinguish regions in wavelength that are driving the
differences coinciding with H, Ca, Fe and Mg absorption lines. The
continuum shape blueward of $4000$\AA\, also has an impact, but in
this plot we are showing only the high frequency signal of the
residuals normalized to the continuum. While traditional indices
based approaches will consider some subset of this data, by
analyzing the full spectrum we are able to determine the metallicity
of individual galaxies more accurately.

\section{The cosmic metallicity history}

In order to combine the metallicity histories of many galaxies at
different redshifts it is necessary to estimate metallicity
histories in a common time frame. In previous work we have formed
complex transform functions to redistribute a galaxies SF in a new
set of bins, but in this case we instead chose a narrow redshift
range of galaxies ($|z-0.1|<0.01$). For our common bins we use the
original time bins (see \citet{PJHC06}) with $t=0$ at $z=0.1$. This
gives a sample of 42160 galaxies.

\begin{figure}
\includegraphics[width=\columnwidth]{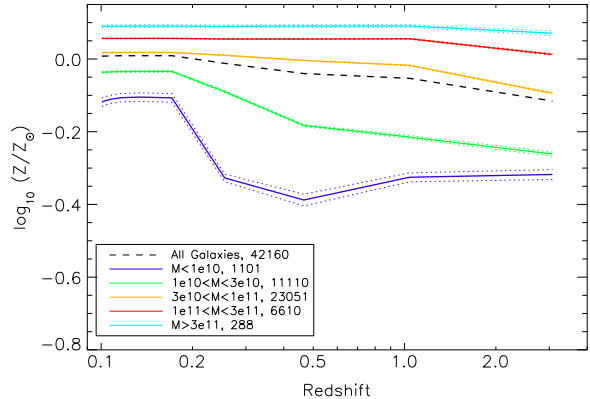}
\caption{Bulk metallicity of stars in SDSS galaxies for different
stellar masses. This is the average metallicity of all stars present
at a particular redshift, regardless of age. To avoid possible
errors from shifting bins we have chosen a narrow redshift range
centred at $z=0.1$. The number of galaxies in each mass range is
given in the inset. Note that more massive galaxies are more metal
rich than less massive galaxies. The enrichment is very flat as a
function of redshift for the more massive galaxies, as the bulk of
their stars formed early and the tiny amount of recent SF has not
been sufficient to change the bulk metallicity. The dotted lines
correspond to a $1\sigma$ bootstrap error.} \label{fig:stars}
\end{figure}

The simplest interpretation of the metallicity history is that of
the bulk of stars in a galaxy at a given time. To calculate this for
each bin we weight the metallicity in the bins up to that point by
the mass of stars remaining from that bin (ie, we include
recycling). This traces the average metallicity of stars in a galaxy
over its evolution, and is shown in Fig. \ref{fig:stars}.

\begin{figure}
\includegraphics[width=\columnwidth]{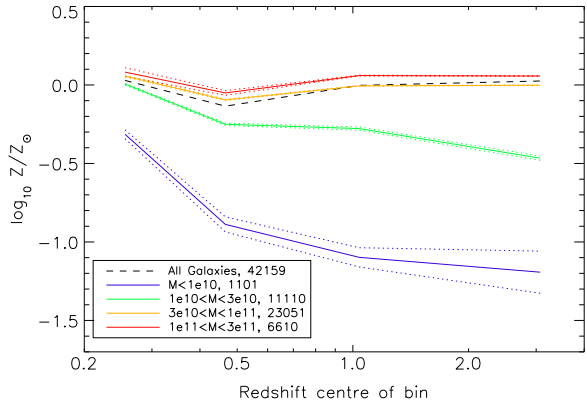}
\caption{Average metallicity for different stellar masses, as in
Fig. 1 except for star forming gas rather than stars.}
\label{fig:massweight}
\end{figure}

The individual metallicity histories of each galaxy can also be
combined to give the mass-weighted cosmic metallicity history of the
star forming gas. To calculate this we average the metallicities for
a given bin, weighted by the mass of that population. We must also
impose a cut to ensure that the metallicity of populations which
have a very small contribution to the spectrum are not considered, as
these have poorly determined metallicity and give results dependent
on the metallicity prior. To avoid this contamination we add a
further cut: to contribute to the average of a particular
time bin a galaxy must have $>25\%$ of its spectral flux in that bin
(see Fig.~\ref{fig:massweight}).

\begin{figure}
\includegraphics[width=\columnwidth]{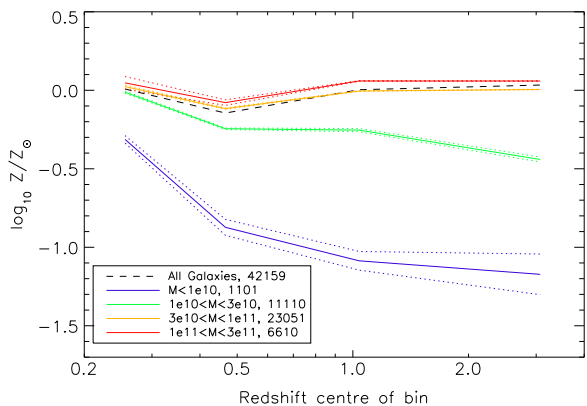}
\caption{Same as Fig.~\ref{fig:massweight} but for a total
fractional spectral flux weighted average.} \label{fig:lightweight}
\end{figure}

Weighting by bin mass obviously biases the results towards high-mass
galaxies. We also present results using a weighting scheme that uses
the fraction of the total spectral flux. Weighting by mass means
that there is a slight bias towards older populations, which may not
be as well determined as for a given mass as the younger populations
give much more light. To calculate the cosmic metallicity history
the metallicities in the bins are then averaged, weighted by the
fractional contribution of the bin to the galaxy luminosity. In this
weighting scheme a galaxy's mass is irrelevant. The corresponding
cosmic metallicity history is shown in Fig.~\ref{fig:lightweight}
and is very similar to the flux-weighted history (Fig.
~\ref{fig:massweight}).

The first thing we learn is that most metals are locked in the most
massive galaxies and that the metallicity evolution for each mass
range is relatively flat as a function of redshift, with variations
of 0.1-0.2 dex for each mass range. Note that a similar trend is
found for a light-weighted plot. Our results compare favourably with
the trends observed at high redshift by \citet{Maiolino07}. If we
look at their Fig.~3 and compare their abundance as a function of
redshift for a stellar mass of $1 \times 10^{10}$ M$_{\odot}$ with
our green line in Fig.~1, 2 and 3 we find that their abundance
values at $z = 3, 0.07$ and $0$ are $\log (Z/Z_{\odot}) = -0.54,
-0.1, 0.1$ respectively, assuming that $12 + \log (O/H)_{\odot} =
8.66$. This compares with our derived values, for the same redshift
intervals, of $-0.5, -0.2, 0.0$, which is within $0.1$ dex of their
values, which is remarkable. Any conclusions are subject to model
choice, as we explore in \S 8.

\section{Mapping the metallicity evolution of the Universe}
\begin{figure*}
\includegraphics[width=2.1\columnwidth]{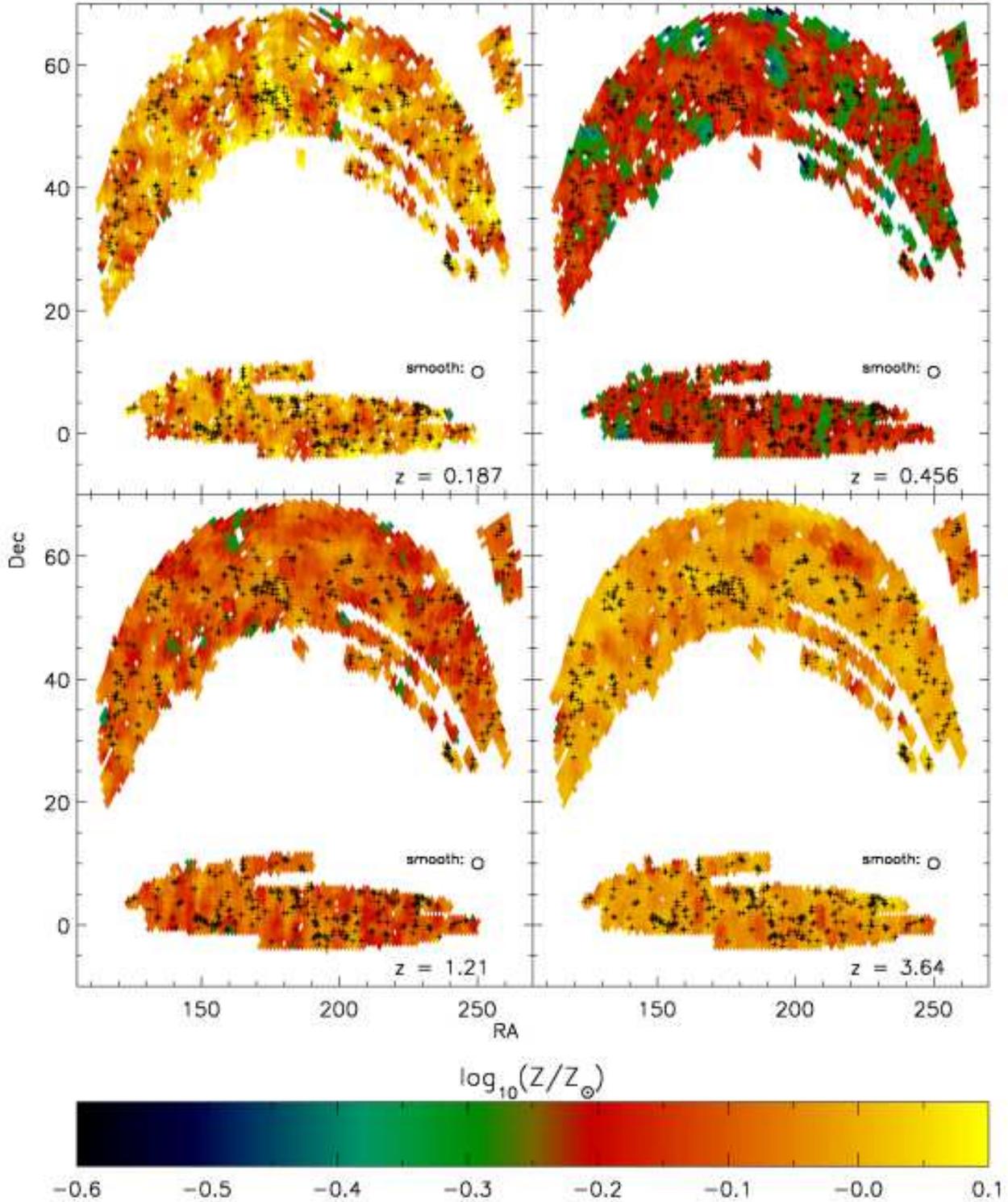}
\caption{HEALPix projection of the mass weighted gas metallicity for
several look back bins averaged over galaxies in the redshift range
$0.0<z<0.1$. Galaxies only contribute to a bin if $>25\%$ of their
star formation occurs in that bin. The overlaid crosses correspond
to the bright cluster members in the C4 catalog. The metallicities
are boxcar smoothed over a $2^{\circ}$ radius. The size of the
smoothing patch is shown in the bottom right of each panel}
\label{fig:maps}
\end{figure*}
The SDSS-DR3 spectroscopic footprint covers $3732$ sq. deg., roughly
10\% of the sky. We use the metallicity history of the galaxies to
create maps over this area of the enrichment history at different
epochs. We use the HEALPix\footnote{Details of the HEALPix package
are available from \tt{http://healpix.jpl.nasa.gov}} algorithm to
determine equal area patches on the sky, and calculate the mass
weighted average metallicities for each patch and time bin as
before. Figure~\ref{fig:maps} shows the mass-weighted metallicity
maps for our four highest redshift bins, smoothed with a boxcar
filter of radius $2^{\circ}$. Over-plotted are the locations of the
Brightest Cluster Galaxies from the SDSS C4 catalog \citep{C4}, used
to represent the distribution of cluster galaxies on the sky. It is
clear by eye that in many regions the crosses follow the regions of
higher metallicity

For areas of the footprint where cluster galaxies exist, a
cross-correlation analysis between mass weighted average gas
metallicity and number of cluster members in cells reveals strong
correlation between the three oldest bins $z=0.456,1.21,3.64$ and
the number of cluster galaxies\footnote{NB: We correlate with
cluster members, not BCGs} (see Tables \ref{table:pcorr} and
\ref{table:scorr}). It would appear that, as stated in
\citet{SJPH06}, metallicity is strongly correlated with environment
- this can be interpreted as the seeds of clusters being the seeds
of metal enrichment in the universe. Note that the overall level of
enrichment in the map at $z=3.64$ is very homogeneous at around the
solar metallicity value, while at $z=0.187$ there is much more
variation. If we assume that metallicity is an indicator of
environment, this offers a tantalisingly glimpse of the growth in
the influence of dark matter structure, only visible by examining
the huge volume at high (temporal rather than spatial) redshift
offered by the fossil record.

 It will be interesting to cross-correlate these maps with
the upcoming SZ experiments which will detect clusters of galaxies
at higher redshift. The larger variations in the lowest redshift map
probably reflect the change in sampling to less massive galaxies, as
these are those in the sample which are likely to have a high
fraction of younger star formation.

\begin{table}
 \centering
 \begin{minipage}{\columnwidth}
  \caption{Pearson correlations of metallicity and number of cluster galaxies in a cell}

\begin{tabular}{| c || c | c | c | c |}
  \hline
  z & \multicolumn{4}{|c|}{Smoothing Radius($^\circ$)}\\
   & 0 & 2 & 3 & 4 \\
\hline
0.187 & 0.091 &  0.118 &  0.104 &  0.079 \\
0.456 & 0.101 &  0.167  & 0.171 &0.171 \\
1.21 & 0.131 &  0.213  & 0.212 & 0.189 \\
3.64 & 0.127 & 0.144 &  0.136  & 0.112 \\

  \hline
  \label{table:pcorr}
\end{tabular}
\end{minipage}
\end{table}

\begin{table}
 \centering
 \begin{minipage}{\columnwidth}
  \caption{Spearman Correlations of metallicity and number of cluster galaxies in a cell}

\begin{tabular}{| c || c | c | c | c |}
  \hline
  z & \multicolumn{4}{|c|}{Smoothing Radius($^\circ$)}\\
  & 0 & 2 & 3 & 4 \\
\hline

0.187 & 0.077 &  0.097 &  0.078 &  0.061 \\
0.456 & 0.084 &  0.179 &  0.182 &  0.167 \\
1.21 & 0.127 & 0.274 &  0.273 &  0.246 \\
3.64 & 0.169 & 0.201 &  0.189 &  0.154 \\

  \hline
  \label{table:scorr}
\end{tabular}
\end{minipage}
\end{table}

\section{The average Mass-metallicity relation}

We now turn our attention to the local mass-metallicity relation for
the SDSS galaxy population to understand its origin and its time
evolution.

\begin{figure}
\includegraphics[width=8.5cm]{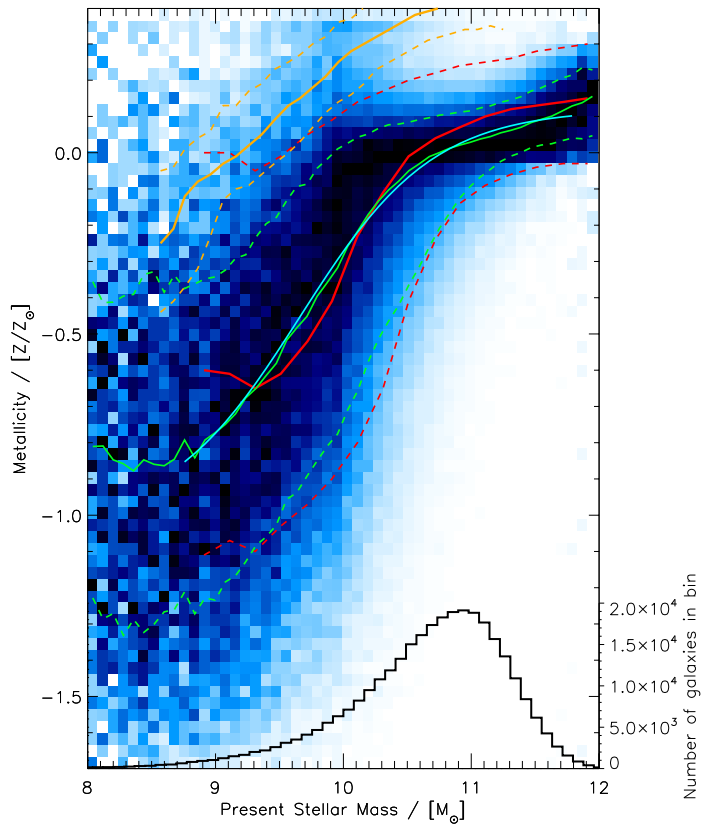}
\caption{Mean metallicity mass relation using the present mass
fractions as weights. The solid green lines represent the 16th, 50th
and 84th percentiles of the distribution and the light blue a
$\tanh$ fit to the results. The dashed and dot-dash lines are the
same percentiles for the \citet{Gallazzi05} (red) and
\citet{Tremonti04} (orange) results. The distributions are
normalized over each column of mass, the underlying galaxy mass
distribution is shown in the lower histogram} \label{fig:massmet}
\end{figure}

Fig.~\ref{fig:massmet} shows the mass-metallicity relation for
$312,815$ galaxies in the DR3 Main Galaxy Sample. In this case we
calculate the mass fraction weighted metallicity, ie for an
individual galaxy the metallicity is calculated by weighting the bin
metallicities by the observed fraction of mass in that bin. The
three solid lines represent the 16th, 50th and 84th percentiles of
the distribution. The first thing to note is that there is a clear
mass-metallicity relation: more massive galaxies harbour stars with
higher metallicity. The average metallicity of an L$_{*}$ galaxy is
solar. For lower masses, the metallicity decreases approximately by
$0.5$dex for every dex in mass.

For galaxies with stellar masses of about $10^9$ M$_{\odot}$, the
average metallicity of the stars is 0.1 the solar value.  Note that
for masses larger than $10^{11}$ M$_{\odot}$ there is a flattening
of the mass-metallicity relation. The maximum value we obtain for
the mass-metallicity relation is $1.1 Z_{\odot}$. The spread in the
relation is also smaller at higher masses ($0.15$ dex) and grows at
smaller masses ($0.5$ dex).

There is a break around $ M_*=10^{10} M_\odot$, below which the dispersion
around the median value increases, and a much wider range of
metallicities is recovered for a given mass. This can be interpreted
as consistent with the turnover point noted in \citet{Kauffmann03,
JPHV05}.

We can obtain a reasonable fit to the median line using a tanh
function over the mass range where cells contain more than 2000
galaxies ($8.8<\log M_*<11.8$) of the form
\begin{equation}
 \log \frac{Z}{Z_\odot}=A+B\tanh \left(\frac{\log M_* - \log
 M_c}{\Delta}\right)
\end{equation}
where $\log M_c=9.66$, $\Delta=1.04$, $A=-0.452$, $B=0.572$ and masses are in solar masses. This
formula is given purely for convenience, and should not be taken to
reflect any underlying physical motivation or be extrapolated
outside of the fitted mass range. It is of interest though that the
break at $\log M_c=9.66$ neatly corresponds with the increase in
dispersion noted above.

\begin{figure}
\includegraphics[width=8.5cm]{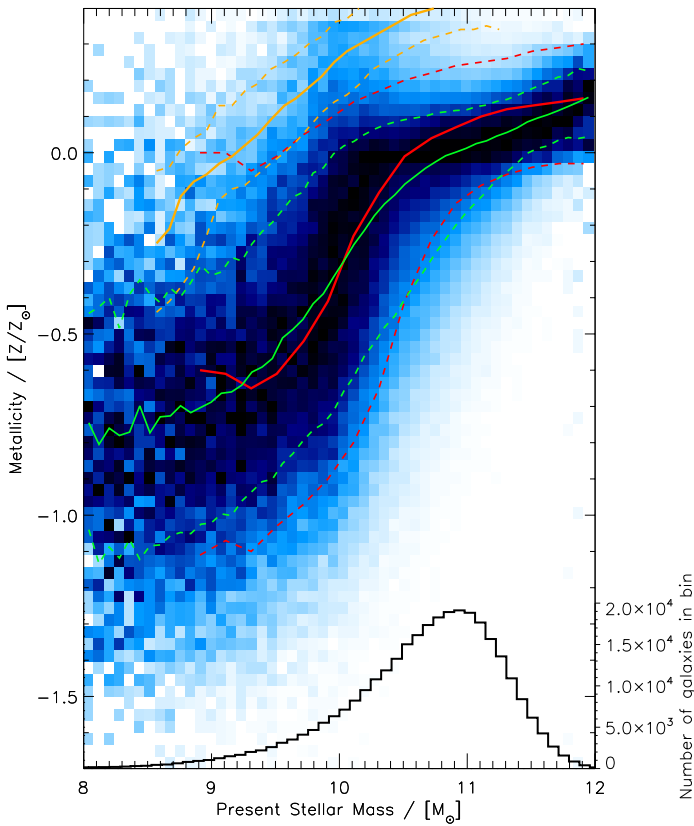}
\caption{Mean metallicity mass relation using the luminous fractions
as weights. The solid line represents the 16th, 50th and 84th
percentiles of the distribution. The dashed and dot-dash lines are
the same percentiles for the \citet{Gallazzi05} (red) and
\citet{Tremonti04} (orange) results. The distributions are
normalized over each column of mass, the underlying galaxy mass
distribution is shown in the lower histogram} \label{fig:masslummet}
\end{figure}

\begin{figure}
\includegraphics[width=\columnwidth]{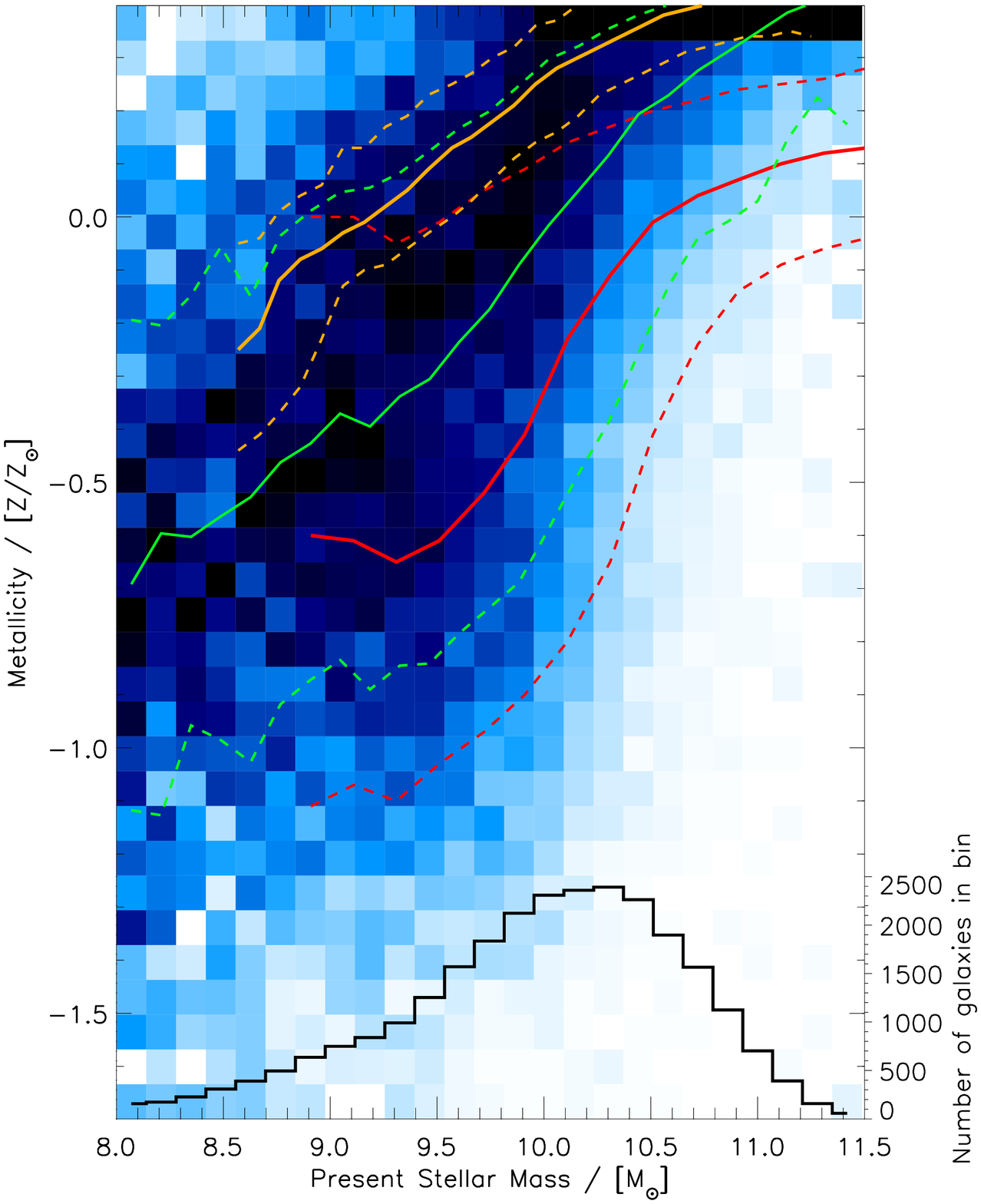}
\caption{Intermediate-age mass-metallicity relation. To compose this
plot we have considered only those galaxies with more than half
their spectral flux in populations $\leq$ 1Gyr. The metallicity is
the average of the younger bins weighted by their fractional
spectral flux. The solid line represents the 16th, 50th and 84th
percentiles of the distribution. The dashed and dot-dash lines are
the same percentiles for the \citet{Gallazzi05} (red) and
\citet{Tremonti04} (orange) results. The distributions are
normalized over each column of mass, the underlying galaxy mass
distribution is shown in the lower histogram} \label{fig:young}
\end{figure}

\begin{figure}
\includegraphics[width=\columnwidth]{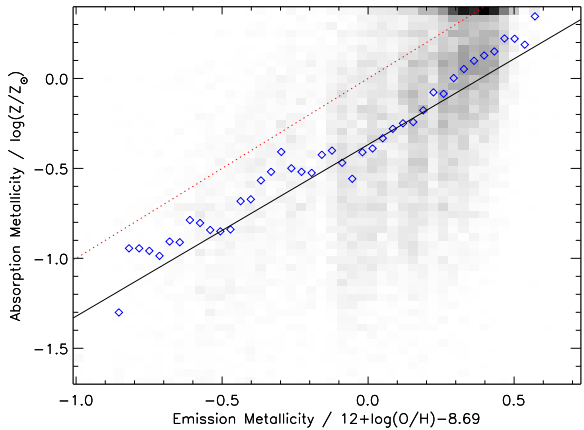}
\caption{A comparison of metallicities determined by absorption and
emission analysis. This plot considers only those 14038 galaxies
with more than half their spectral flux in populations $\leq$ 1Gyr
that also have a metallicity reported in the \citet{Tremonti04}
sample. The absorption metallicity is the average of the younger
bins weighted by their fractional spectral flux and the emission
metallicity is taken from \citet{Tremonti04}. The blue diamonds show
the median value of absorption metallicity for a given emission
metallicity, the black line is a linear fit through all points and
the red dotted line corresponds to a 1:1 relation.  See text for a discussion of the offset.}
\label{fig:tremonti}
\end{figure}

\begin{figure}
\includegraphics[width=\columnwidth]{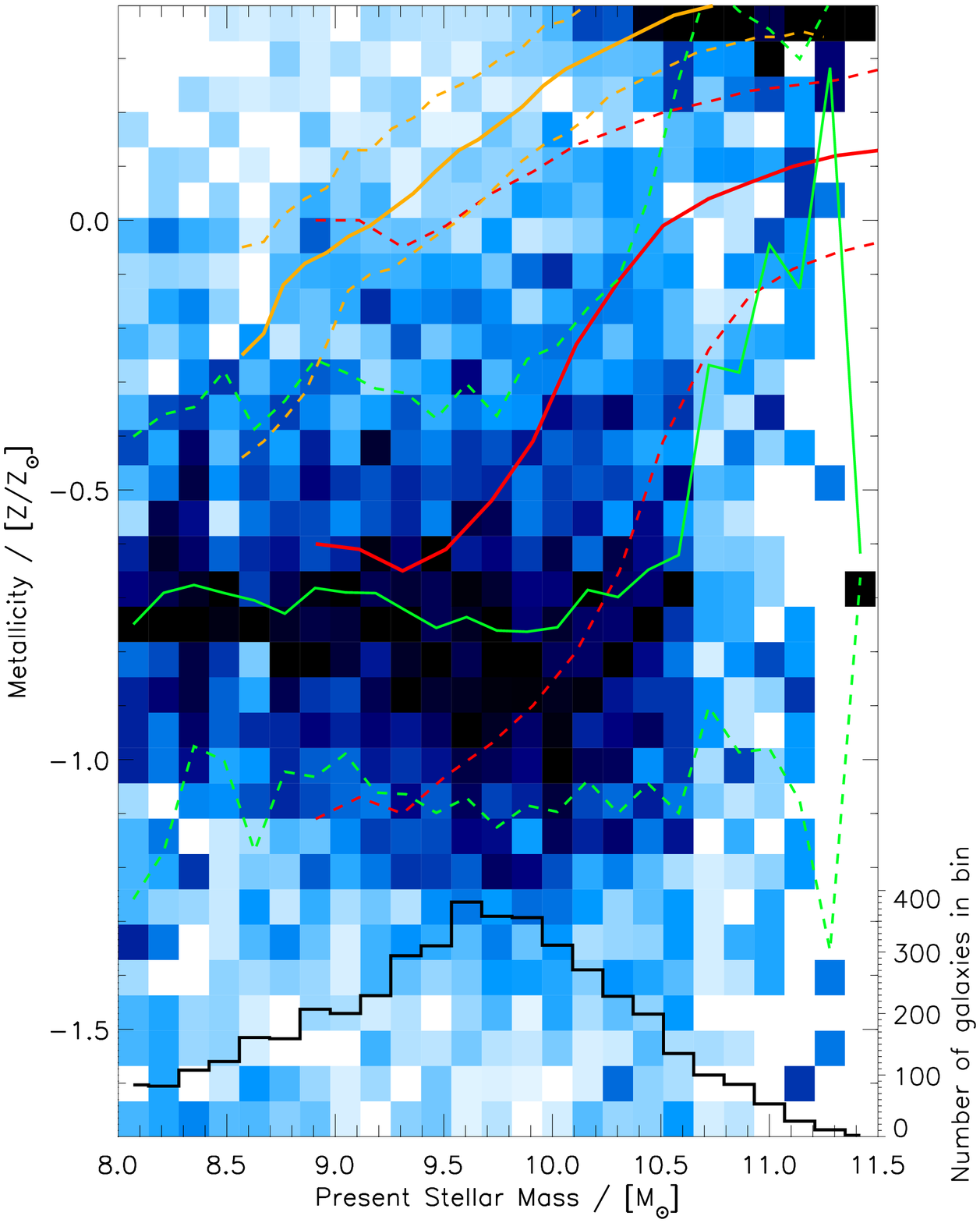}
\caption{Youngest  age mass-metallicity relation. To compose these
plots we have considered only those galaxies with more than half
their spectral light in populations $<$ 1Gyr. The metallicity is the
average of the younger bins weighted by their fractional spectral
flux. The solid line represents the 16th, 50th and 84th percentiles
of the distribution. The dashed and dot-dash lines are the same
percentiles for the \citet{Gallazzi05} (red) and \citet{Tremonti04}
(orange) results. The distributions are normalized over each column
of mass, the underlying galaxy mass distribution is shown in the
lower histogram} \label{fig:v_young}
\end{figure}

We can now compare this relation with previous work. Recently
\citet{Gallazzi05}, using absorption indices in the spectra of SDSS
galaxies, have determined the local mass-metallicity relation. Their
approach is different from ours since they concentrate on specific
absorption features of the  spectrum, and their results are weighted towards the most luminous populations. For most old
galaxies this will be very similar to our own result, but because we
know the contribution to the spectra of each component we can
recalculate the average metallicity of galaxies with similar
weighting (Fig. \ref{fig:masslummet}). The agreement is very good
over the matching mass range, although seems to be breaking down at
the low-mass end. This is not surprising because due to downsizing
of the galaxy population, it is the low-mass galaxies that dominate
star formation today. This of course is due to the fact that for
younger galaxies with ongoing star formation it is the younger stars
that are more luminous, even though by mass the galaxies are almost
certainly dominated by older stars \citep{PJHC06}. The otherwise
excellent agreement reinforces the fact that we are not introducing
degeneracies in the overall quantities by recovering the star and
metallicity history in $11$ time bins. In Fig.~\ref{fig:masslummet}
we show the same plot as in Fig.~\ref{fig:massmet} but this time
luminosity-weighted (strictly, by spectral flux). Note that in this
case there is again good agreement, slightly better than before for
lower masses.

We have also overplotted the mass-metallicity relation determined by
\citet{Tremonti04} from emission line measurements, and a clear
offset can seen. In the next section we take advantage of MOPED
ability to recover the time evolution of the metallicity and star
formation history to explain the difference.

\section{Recovering the mass-metallicity for younger populations}

Since the MOPED algorithm recovers both the star formation and
metallicity history of a galaxy it can be used to investigate the
evolution of the metallicity for a given galaxy - a task impossible
with simple indices. In particular we can look at only galaxies with
recent star formation and examine how the metallicity of the star
forming gas has evolved from that which produced the majority of the
stars at earlier ages.

By using only the younger populations ($\leq 1$Gyr) of galaxies for
the analysis we can reconstruct the information that would be probed
by the emission line analysis (as in \citet{Tremonti04}). Note
that our analysis explicitly removes strong emission lines - all our
information comes from continuum shape and absorption features.
Excluding AGN activity, the emission lines reflect the metallicity
of nebular gas around star-forming regions that has been excited by
the UV emission of young stars. Fig.~\ref{fig:young} shows the
mass-metallicity relation of the star-forming gas as derived
indirectly from our analysis. It is important to use only
metallicities which are well determined - for this analysis we use
the 27915 galaxies for which over half the spectral light comes from
these younger populations, and plot the metallicity weighted by
their fractional spectral flux. Note that now the agreement with the
\citet{Tremonti04} result is much better and our analysis shows that
their metallicities are dominated by an intermediate $\sim 1$ Gyr
stellar population in sub$-L^{*}$ galaxies. In Fig.
\ref{fig:tremonti} we extend this analysis by comparing directly the
metallicities determined from absorption and emission of galaxies
which have more than half of their flux in populations $\leq 1$Gyr.
The figure shows good agreement, confirming that given sufficient
flux in those populations the MOPED algorithm using absorption
features can determine accurate metallicities. The offset could
indicate a systematic difference in metallicity between the birth
cloud (responsible for the emission features) and the young stars,
or possibly a calibration offset in the models.

We can extend this analysis to galaxies younger than the 1 Gyr bin,
although the number of galaxies with sufficient star formation in
these youngest age bins is very small and determination of
metallicity in these younger stars with weaker absorption features
is more difficult than in older, cooler stars.
Fig.~\ref{fig:v_young} shows the mass-metallicity relation of these
younger populations and it is clear that any relationship has broken
down. First, the stellar mass range is much smaller than before
because of downsizing in the galaxy population. Only galaxies with
stellar masses below $10^{10}$ M$_{\odot}$ contribute significantly
to the young population metallicity and the mass range is only of
$1.5$ dex. As pointed out earlier the mass-metallicity relation is
flat (at least for $M<10^{10.5}M_{\odot}$), with no dependence on
mass and an average metallicity of $0.2 Z/Z_{\odot}$. Why galaxies
below this mass all have the same metallicity is something we will
investigate in future work, but one could envision environment as a
major driver, where the gas in filaments that feed the galaxy has
been enriched to a certain level.

\section{Model dependence of the Metallicity History}

\begin{figure}
\includegraphics[width=\columnwidth]{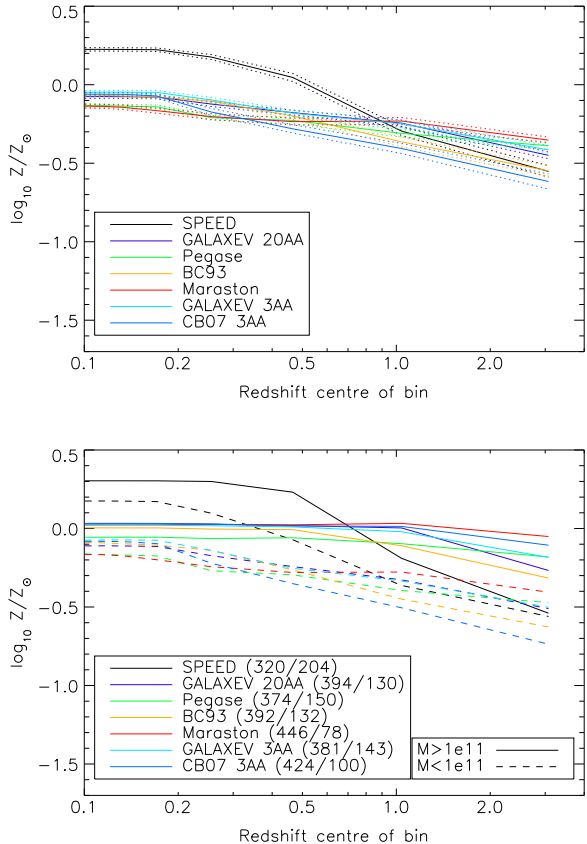}
\caption{Cumulative metallicity of stars in SDSS galaxies recovered
using different stellar population models. The lower panel
illustrates the difference between low and high mass galaxies.}
\label{fig:comp_b}
\end{figure}

The results shown so far have all been generated using the current
standard in stellar modelling codes, \citet{BC03}. It is important
however to consider the effect of changing the models used; we make
no statement as to which set is ``right", and emphasize that the set
of galaxies chosen for comparison, detailed in \citet{PJHC06}, is
relatively small (808).

We have investigated the impact of model choices by recovering the
metallicity history with a suite of different stellar population
models: SPEED \citep{Jimenez+04}, PEGASE \citep{pegase}, BC93
\citep{BC93}, Maraston \citep{maraston05}, GALAXEV \citep{BC03},
CB07 (As BC03 but an with improved treatment of TP-AGB stars). Our
reference model is the BC03 at a spectral resolution of 3\AA.

In Fig.~\ref{fig:comp_b} we show the differences in average
metallicity of stars as a function of redshift. It is clear that our
conclusions regarding higher-mass galaxies having higher metallicity
is robust whichever model is used. The deviation shown by the SPEED
models reflects the fact that, as shown in \citet{PJHC06}, the star
formation history recovered by this model is broader, with overall
SFR peaking at $z<1$. It should also be noted that the SPEED models
allow a wider range of metallicity values than other synthesis
codes.

\begin{figure}
\includegraphics[width=\columnwidth]{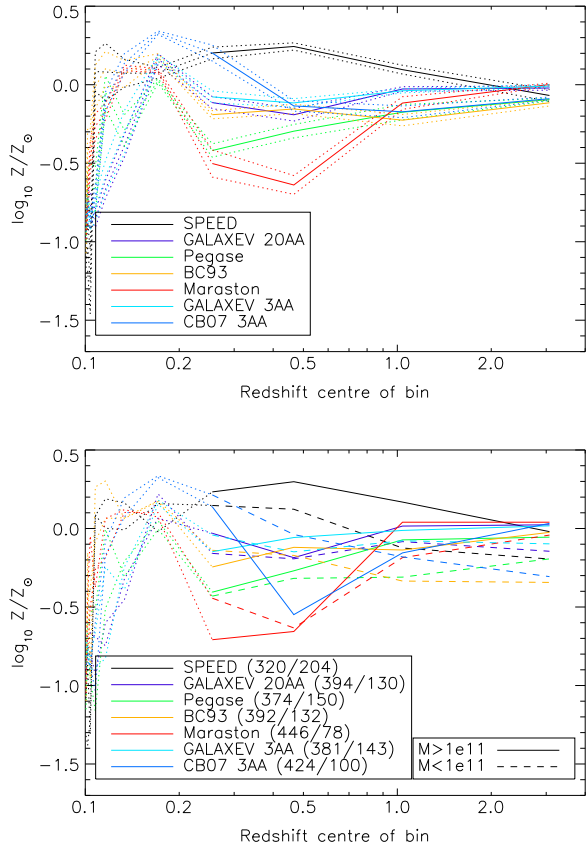}
\caption{Metallicity of star forming gas in SDSS galaxies recovered
using different stellar population models. We caution that the
region lower than $z=0.3$ (dotted) has relatively little flux, and
is not used for our earlier analysis.} \label{fig:comp}
\end{figure}

In Fig.~\ref{fig:comp}, there is apparently some issue at z=0.2 ($t\sim1$ Gyr), where almost all the models give very similar
metallicity. With a population of galaxies taken at $z=0.1$, this
corresponds to a population of stars of age 1 Gyr, in some respects
a difficult population to identify as no part of the spectrum is
dominated by this population for usual star formation histories
\citep{PJHC06, Mathis06}. As this age reflects the point for the
average galaxy where flux per population is at a minimum, we caution
against over-interpretation. Although it may be a systematic failing
of the libraries used to calibrate the various sets of models it
seems more likely that what is being reflected is the prior
introduced by a fiducial metallicity. Alternatively, it could be
that a more sophisticated treatment of the post-AGB stages, as
included in the more modern models, removes the problem. Whichever
set of models is chosen, the redshift evolution of the stellar
metallicity in the SDSS population is fairly flat, with a variation
of about 0.5 dex.

Of particular interest are the differences between the Maraston,
SPEED and CB07 models. These three each include a treatment of mass
loss along the thermal-pulsating AGB phase of stellar evolution. The
SPEED models incorporate this effect using an empirical prescription
\citep{Jorgensen91,Jimenez+98,Jimenez+04}. If we focus on these
models which include mass loss (Maraston, SPEED and CB07) then the
discrepancy in recovery of metallicity is of about $0.1$ dex beyond
z=1, but diverges by 0.5 dex at lower redshifts. Clearly the
metallicity is more sensitive to the model choice than the star
formation history (see \citet{PJHC06}) and the error budget in
determining the metallicity history is dominated by the choice of
model. Further work is needed to calibrate the models with higher
accuracy.

\begin{figure*}
\includegraphics[width=1.75\columnwidth]{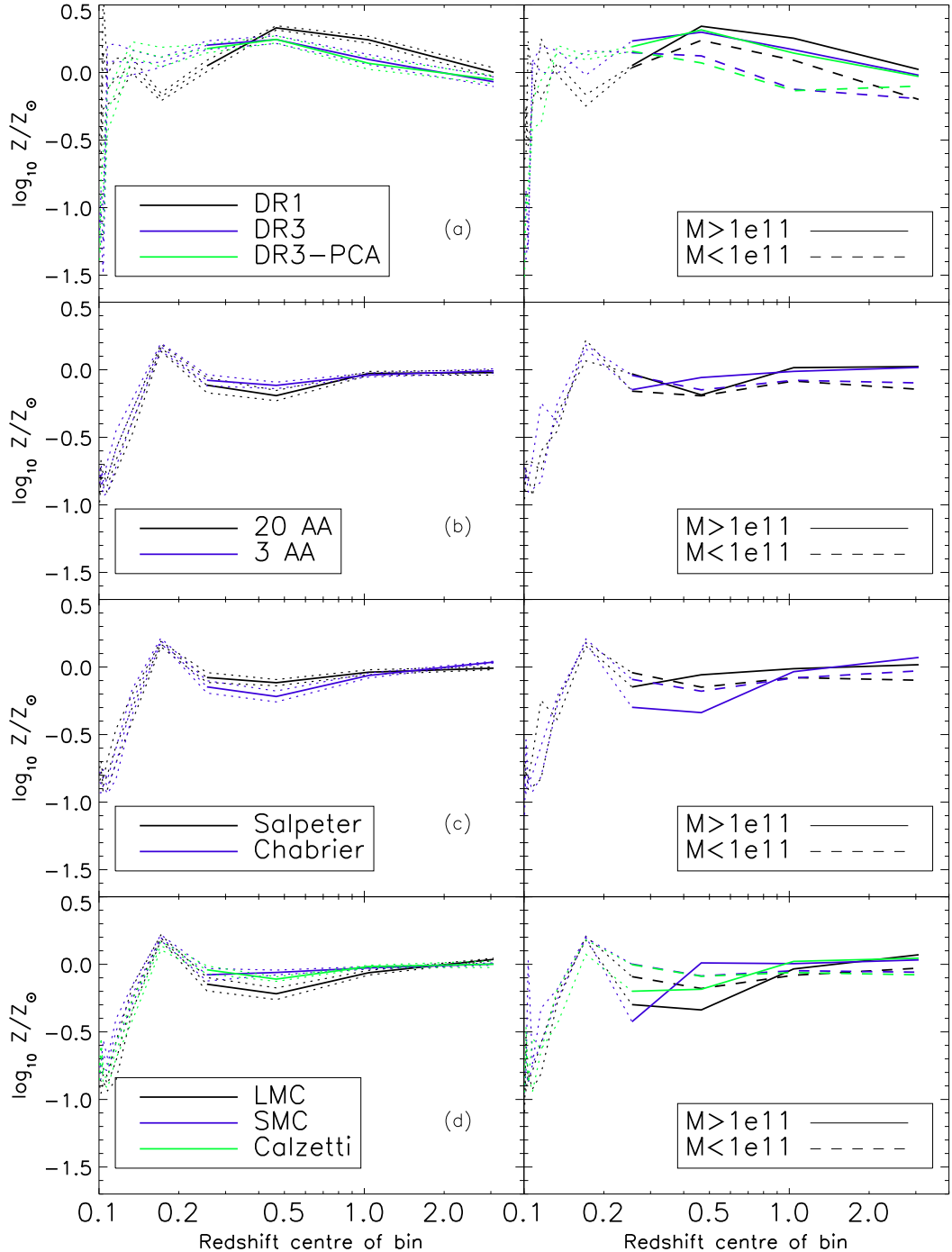}
\caption{Effect of several systematic effects on the determination
of the metallicity from the spectra of the stellar populations of
galaxies (left), and how they effect conclusions based on mass
(right). The different panels show the effect of different
systematics: Fig. (a) shows the effect of different pipelines in the
data reduction of the SDSS spectra, the error is comparable to the
random error, although it is clear that there is a difference
between the DR1 and DR3 reductions. Fig. (b) shows the effect of
spectral resolution in the models; it would not be surprising to see
a difference when the spectral resolution is increased, but this is
not evident. This could be due to the fact that the underlying model
(BC03) is calibrated at 3\AA\ and rebinned to 20\AA. Fig. (c) shows
the difference by choice of IMF - the overall change in recovered
metallicity of the star forming gas is small, but the large changes
in mass around the cut (due to the strong dependence of mass on the
IMF) give a misleading effect on the right hand plot. Fig. (d) shows
the dependence on dust model choice, and again the variation on the
right hand figure is due to mass differences rather than a large
difference in recovered metallicity.} \label{fig:changes}
\end{figure*}

Figure~\ref{fig:changes} shows in different panels the effect of
various systematics. It is clear from Fig. \ref{fig:changes}(a) that
changing the photometric reduction pipeline will have an effect on
the metallicity values recovered (the SPEED model was used for this
comparison, GALAXEV for the following). The changes from DR1 to DR3
are clear, but it appears that PCA skyline \citep{Wild05} cleaning
does nothing to change the recovered metallicity history. Fig.
\ref{fig:changes}(b) makes it clear that, for a model calculated and
calibrated at 3\AA\ but rebinned to 20\AA\, it is still possible to
recover the same metallicity history. It is interesting however to
note that this is the only change which has any effect on the 1 Gyr
bump evident in most of other stellar population models, but further
analysis is required (with many more galaxies and rebinning
resolutions) to resolve this issue. Figures \ref{fig:changes}(c) and
\ref{fig:changes}(d) show that changing IMF and dust extinction
curve do not alter the metallicities recovered by a large amount.
The apparently large variations in the right hand panels can be
explained by the changes in galaxy mass, caused by changing either
the IMF or extinction curve, moving galaxies between the two
samples.

\section{Conclusions}
In this paper we have presented the cosmic evolution of the bulk
metallicity of galaxies, demonstrating a clear trend of downsizing
which is robust to the selection of stellar population model to
determine the metallicity of the galaxy. We have extended this to
explore the evolution of metallicity of star forming gas, which also
follows a downsizing scenario but is more dependent on the choice of
stellar model. The trend shows how massive galaxies have higher
metallicities at earlier times while less massive galaxies have
lower metallicities at earlier times but reach the same metallicity
at about 1 Gyr from the current time. We have shown a clear
correlation between metallicity and cluster environment, and will
perform more correlations as further datasets become available.

We have reconciled the disparity between the mass-metallicity
relations recovered from nebular emission line spectra and stellar
absorption methods. As one might expect, the absorption method is
entirely appropriate for estimating the metallicity of older,
established galaxies, while it may fail for galaxies which are
undergoing recent star formation where the most luminous population
is almost certainly not the most massive in stellar mass. We have
verified that the fossil analysis technique, which here excludes
emission lines, can be used simultaneously to uncover the
metallicities of both young and old populations by producing a mass-metallicity relation that accurately maps the results given by both
emission and absorption line diagnostics.

It is clear from the comparison of stellar models that the
metallicity history determined for the star forming gas is highly
model dependent. Although we use models with a wide range of
publication dates, it is apparent that the three most modern show
the largest deviation. We acknowledge that alpha enhancement, not
included in any of the stellar population models we consider, could
affect the metallicities we recover.

\section*{acknowledgments}

A portion of BDP's work was supported by the Alexander von Humboldt
Foundation, the Federal Ministry of Education and Research, and the
Programme for Investment in the Future (ZIP) of the German
Government. We acknowledge the use of HEALPix \citep{gorski} and
software packages developed by David Fanning (Fanning Consulting),
and the assistance of Gerard Lemson and GAVO for access to the
Millennium/SDSS Database and advice on SQL. We thank Manuchehr
Taghizadeh-Popp for assistance in the identification of SDSS
galaxies in the Healpix tessellation scheme. Funding for the SDSS
has been provided by the Alfred P. Sloan Foundation, the
Participating Institutions, the National Science Foundation, the
U.S. Department of Energy, NASA, the Japanese Monbukagakusho, and
the Max Planck Society. RJ is supported by grants from the Spanish
Ministry of Science (MEC), the Spanish Higher Council for Scientific
Research (CSIC) and the European Union (IRG).


\begin{thebibliography}{99}
\bibitem[Abazajian et al.(2005)]{SDSS-DR3} Abazajian, K., et al.
         (the SDSS collaboration)\ 2005, AJ, 129, 1755

\bibitem[\protect\citeauthoryear{Bouch{\'e} et
al.}{2007}]{Bouchet07} Bouch{\'e} N., Lehnert M.~D., Aguirre A.,
P{\'e}roux C., Bergeron J., 2007, MNRAS, 378, 525

\bibitem[\protect\citeauthoryear{Bruzual \& Charlot}{1993}]{BC93} Bruzual A.~G., Charlot S., 1993, ApJ, 405, 538

\bibitem[\protect\citeauthoryear{Bruzual \& Charlot}{2003}]{BC03} Bruzual G., Charlot S., 2003, MNRAS, 344, 1000


\bibitem[\protect\citeauthoryear{Cid Fernandes et~al.}{2005}]{Sodre05}
{Cid Fernandes},  R., {Mateus}, A., {Sodr{\'e}}, L.,{Stasi{\'n}ska},
G., \& {Gomes}, J.~M.\ 2005, MNRAS, 358, 363


\bibitem[\protect\citeauthoryear{Cid Fernandes et al.}{2007}]{Cid07} Cid Fernandes R., Asari N.~V., Sodr{\'e} L., Stasi{\'n}ska G., Mateus A., Torres-Papaqui J.~P., Schoenell W.,
2007, MNRAS, 375, L16

\bibitem[\protect\citeauthoryear{Erb}{2008}]{Erb08} Erb
D.~K., 2008, ApJ, 674, 151


\bibitem[\protect\citeauthoryear{Erb et al.}{2006}]{Erb06}
Erb D.~K., Shapley A.~E., Pettini M., Steidel C.~C., Reddy N.~A.,
Adelberger K.~L., 2006, ApJ, 644, 813

\bibitem[\protect\citeauthoryear{Fioc \& Rocca-Volmerange}{1997}]{pegase} Fioc M., Rocca-Volmerange B., 1997, A\&A, 326, 950

\bibitem[\protect\citeauthoryear{Freeman \&
Bland-Hawthorn}{2002}]{FB06} Freeman K., Bland-Hawthorn J., 2002,
ARA\&A, 40, 487

\bibitem[Gallart et
al.(2005)]{Gallart05} Gallart, C., Zoccali, M., \& Aparicio, A.\
2005, AR\&A, 43, 387


\bibitem[Gallazzi et al.(2005)]{Gallazzi05} Gallazzi, A., Charlot,
S., Brinchmann, J., White, S.~D.~M., \& Tremonti, C.~A.\ 2005,
MNRAS, 362, 41

\bibitem[Gallazzi et al.(2006)]{Gallazzi06} Gallazzi, A., Charlot,
S., Brinchmann, J., \& White, S.~D.~M.\ 2006, MNRAS, 370, 1106

\bibitem[\protect\citeauthoryear{Gallazzi et
al.}{2008}]{Gallazi08} Gallazzi A., Brinchmann J., Charlot S.,
White S.~D.~M., 2008, MNRAS, 383, 1439

\bibitem[\protect\citeauthoryear{Gordon et al.}{2003}]{Gordon03} Gordon, K. D., Clayton, G. C., Misselt, K. A., Landolt, A. U., Wolff, M. J., 2003, ApJ, 594, 279

\bibitem[G{\'o}rski et al.(2005)]{gorski} G{\'o}rski, K.~M.,
Hivon, E., Banday, A.~J., Wandelt, B.~D., Hansen, F.~K., Reinecke,
M., \& Bartelmann, M.\ 2005, ApJ, 622, 759

\bibitem[Heavens et al.(2000)]{HJL00} Heavens, A.~F.,
Jimenez, R., \& Lahav, O.\ 2000, MNRAS, 317, 965

\bibitem[Heavens et al.(2004)]{HPJD04} Heavens, A., Panter, B.,
Jimenez, R., \& Dunlop, J.\ 2004, Nature, 428, 625

\bibitem[Jimenez et al.(1998)]{Jimenez+98} Jimenez, R., Padoan,
P., Matteucci, F., \& Heavens, A.~F.\ 1998, MNRAS, 299, 123

\bibitem[\protect\citeauthoryear{Jimenez et~al.}{2004}]{Jimenez+04} {Jimenez} R.,  {MacDonald} J.,  {Dunlop} J.~S.,  {Padoan} P.,    {Peacock}  J.~A.,  2004, MNRAS, 349, 240

\bibitem[Jimenez et al.(2005)]{JPHV05} Jimenez, R., Panter, B.,
         Heavens, A.~F., \& Verde, L.\ 2005, MNRAS, 356, 495

\bibitem[Jorgensen(1991)]{Jorgensen91} Jorgensen, U.~G.\ 1991, A\&A, 246, 118

\bibitem[Kauffmann et al. (2003)]{Kauffmann03} Kauffmann, G. et al.
2003, MNRAS, 341, 33

\bibitem[Kewley \& Dopita(2002)]{Kewley02} Kewley, L.~J., \& Dopita,
M.~A.\ 2002, ApJS, 142, 35

\bibitem[Maiolino et al.(2007)]{Maiolino07} Maiolino, R., et al., astro-ph, arXiv.org:0712.2880

\bibitem[\protect\citeauthoryear{Maraston}{2005}]{maraston05}Maraston C., 2005, MNRAS, 362, 799

\bibitem[Mateus et al.(2006)]{SEAGal06}  Mateus, A., Sodre, L., Cid Fernandes, R., \& Stasinska, G. 2006, MNRAS, submitted (astro-ph/0604063)

\bibitem[Mathis et~al.(2006)]{Mathis06}
    Mathis, H., Charlot, S., \& Brinchmann, J., 2006, MNRAS, 365, 385

\bibitem[Miller et al.(2005)]{C4} Miller, C.~J., et al.\
2005, AJ, 130, 968

 \bibitem[Ocvirk et~al.(2006)]{Ocvirk06}
    Ocvirk, P.,  Pichon, C.,  Lan{\c c}on, A., \& Thi{\'e}baut, E.,
    2006, MNRAS, 365, 46

\bibitem[Panter et al.(2003)]{PHJ03}
    Panter, B., Heavens, A.~F., \& Jimenez, R.\ 2003, MNRAS, 343, 1145

\bibitem[Panter et al.(2004)]{PHJ04}
    Panter, B., Heavens, A.~F., \& Jimenez, R.\ 2004, MNRAS, 355, 764

\bibitem[Panter et al.(2007)]{PJHC06} Panter, B., Jimenez, R.,
     Heavens, A.~F., \& Charlot, S.\ 2007, MNRAS, 378, 1550

\bibitem[Pettini(2001)]{Pettini01} Pettini, M.\ 2001, The Promise
of the Herschel Space Observatory, 460, 113

 \bibitem[Reichardt et al.(2001)]{Reichardt01} Reichardt, C.,
Jimenez, R., \& Heavens, A.~F.\ 2001, MNRAS, 327, 849

\bibitem[Sheth et al.(2006)]{SJPH06} Sheth, R.~K., Jimenez,
R., Panter, B., \& Heavens, A.~F.\ 2006, ApJL, 650, L25

\bibitem[Tojeiro et al.(2007)]{VESPA} Tojeiro, R., Heavens,
A.~F., Jimenez, R., \& Panter, B.\ 2007, MNRAS, 381, 1252

\bibitem[\protect\citeauthoryear{Tremonti et al.}{2004}]{Tremonti04}{Tremonti} C.~A. et al.,  2004, ApJ, 613, 898


\bibitem[\protect\citeauthoryear{Wild \& Hewett}{2005}]{Wild05} Wild V., Hewett P.~C., 2005, MNRAS, 358, 1083


\bibitem[Worthey(1994)]{W94} Worthey, G.\ 1994, ApJS, 95,
107


\end{thebibliography}
\end{document}